\documentclass[conference]{IEEEtran}
\usepackage{amsmath}
\usepackage{latexsym}
\usepackage{graphicx}
\usepackage{bbding}
\usepackage{indentfirst}
\IEEEoverridecommandlockouts
\begin{document}

\title{Optimal Power Allocation for A Massive MIMO Relay Aided Secure Communication}
\author{\authorblockN{Jian~Chen$^{\dagger}$, Xiaoming~Chen$^{\dagger,\ddagger}$, and
Wolfgang~Gerstacker$^{\ddagger}$
\\$^{\dagger}$College of Electronic and Information Engineering, Nanjing
University of Aeronautics and Astronautics, China.
\\$^{\ddagger}$Institute for Digital Communications, Friedrich-Alexander University
Erlangen-N\"{u}rnberg, Germany.}}
\maketitle

\begin{abstract}
In this paper, we address the problem of optimal power allocation
at the relay in two-hop secure communications under practical
conditions. To guarantee secure communication during the
long-distance transmission, the massive MIMO (M-MIMO) relaying
techniques are explored to significantly enhance wireless
security. The focus of this paper is on the analysis and design of
optimal power assignment for a decode-and-forward (DF) M-MIMO relay, so as
to maximize the secrecy outage capacity and minimize the
interception probability, respectively. Our study reveals the
condition for a nonnegative the secrecy outage capacity, obtains
closed-form expressions for optimal power, and presents the
asymptotic characteristics of secrecy performance. Finally,
simulation results validate the effectiveness of the proposed
schemes.
\end{abstract}

\section{Introduction}
The open nature of the wireless channel facilitates a multiuser
transmission, but also incurs the security problem. Recently, as a
complement of traditional upper-layer encryption techniques,
physical layer security (PHY-security) has been proposed to realize
secure communications by making use of the characteristics of
wireless channels, i.e., noise, fading and interference
\cite{Wyner}.

From an information-theoretic viewpoint, the performance of
PHY-security is determined by the rate difference between the
legitimate channel and the eavesdropper channel \cite{SC1}
\cite{SC2}. Therefore, to enhance wireless security, it makes
sense to simultaneously increase the legitimate channel rate and
decrease the eavesdropper channel rate. Inspired by this, various
physical layer techniques have been introduced to improve the secrecy
performance. Wherein, MIMO relaying technique gains considerable
attention due to the following two reasons. First, the relay can
provide a diversity gain and shorten the accessing distance, and
thus improve the secrecy performance \cite{Relay1}. Second, MIMO
techniques, such as spatial beamforming, can reduce the
information leakage to the eavesdropper \cite{Relay2}. The beamforming schemes
at the MIMO relay based on amplify-and-forward (AF) and
decode-and-forward (DF) relaying protocols were presented in
\cite{AF} and \cite{DF}, respectively. It is worth pointing out
that the optimal beam design at the relay requires global channel
state information (CSI) \cite{Beamforming}. Yet, the CSI,
especially eavesdropper CSI is difficult to obtain, since the
eavesdropper is usually passive and keeps silent. Therefore, it is
impossible to realize absolutely secure communications over fading
channels. In this context, statistical secrecy performance metrics,
e.g., secrecy outage capacity and interception probability, are
adopted to evaluate wireless security
\cite{secrecyoutagecapacity}.

Recently, an advanced MIMO relaying technology, namely massive MIMO (M-MIMO)
relaying, is introduced into secure communications to further
improve the secrecy performance \cite{LS-MIMORelaying}. Even
without eavesdropper CSI, M-MIMO relaying can generate a very
high-resolution spatial beam, and thus the information leakage to
the eavesdropper is quite small. More importantly, the secrecy
performance can be enhanced by simply adding more antennas. Hence,
the challenging issue of short-distance interception in secure
communications can be well solved. Note that in two-hop secure
communications, the transmit power at the relay has a great impact
on the secrecy performance, since it affects the signal quality at
both the destination and the eavesdropper. An optimal power
allocation scheme for a multi-carrier two-hop single-antenna
relaying network was given in \cite{PowerAllocation} by maximizing the sum secrecy rate. However, the power allocation for a
multi-antenna relay, especially an M-MIMO relay, is still an open
issue. In this paper, we focus on power allocation for DF M-MIMO
secure relaying systems under very practical assumptions, i.e., no
eavesdropper CSI and imperfect legitimate CSI. The contributions
of this paper are three-fold:

\begin{enumerate}
\item We reveal the relation between the secrecy outage capacity and
a defined relative distance-dependent path loss, and then derive
the condition for a nonnegative secrecy outage capacity
and the constraint on the minimum number of antennas.

\item We derive closed-form expressions for the optimal power at the relay
in the sense of maximizing the secrecy outage capacity and
minimizing the interception probability, respectively.

\item We present clear insights into the secrecy performance
through asymptotic analysis. We show that the maximum secrecy outage
capacity is an increasing function of the source power while the
minimum interception probability keeps constant.
\end{enumerate}

The rest of this paper is organized as follows. We first give an
overview of the DF M-MIMO secure relaying system in Section II,
and then derive two optimal power allocation schemes for the
relay in Section III. In Section IV, we present simulation results
to validate the effectiveness of the proposed schemes. Finally, we
conclude the paper in Section V.

\section{System Model}

We consider a time division duplex (TDD) two-hop massive MIMO
(M-MIMO) relaying system, where a single antenna source transmits
message to a single antenna destination with the aid of a relay
with $N_R$ antennas, while a passive single antenna eavesdropper
intends to intercept the information. Note that the number of
antennas at the relay in such an M-MIMO relaying system is quite
large, i.e., $N_R=100$ or even bigger. It is assumed that there is
no direct transmission between the source and the destination due
to a long propagation distance. The relay system works in a
half-duplex mode, which means any successful transmission requires
two time slots. Specifically, the source sends the signal to the
relay in the first time slot, and then the relay forwards the
post-processing signal to the destination in the second time slot.
In this paper, we assume the eavesdropper is far from the source
and close to the relay, since it was concluded that the signal comes from the
relay directly. Therefore, it is reasonable to assume that the
eavesdropper only monitors the transmission from the relay to the
destination.

We use $\sqrt{\alpha_{S,R}}\textbf{h}_{S,R}$,
$\sqrt{\alpha_{R,D}}\textbf{h}_{R,D}$ and $\sqrt{\alpha_
{R,E}}\textbf{h}_{R,E}$ to represent the channels from the source
to the relay, the relay to the destination, and the relay to the
eavesdropper, respectively, where $\alpha_{S,R}$, $\alpha_{R,D}$
and $\alpha_{R,E}$ are the distance-dependent path losses and
$\textbf{h}_{S,R}$, $\textbf{h}_{R,D}$, and $\textbf{h}_{R,E}$ denote the
channel fading coefficient vectors with independent and
identically distributed (i.i.d.) zero mean and unit variance
complex Gaussian entries. It is assumed that the channels remain
constant during a time slot and fade independently over slots.
Thus, the received signal at the relay in the first time slot can
be expressed as
\begin{equation}
\textbf{y}_R=\sqrt{P_S\alpha_{S,R}}\textbf{h}_{S,R}s+\textbf{n}_R,\label{eqn1}
\end{equation}
where $s$ is the normalized Gaussian distributed transmit symbol,
$P_S$ is the transmit power at the source, and $\textbf{n}_R$ stands for the
additive white Gaussian noise with unit variance at the relay. We
assume that the relay has perfect CSI about $\textbf{h}_{S,R}$ by
channel estimation. Then, maximum ratio combination (MRC) decoding
is performed to recover the information. Specifically, the received
signal is multiplied by a vector
$\textbf{w}_R=\textbf{h}_{S,R}^H/\|\textbf{h}_{S,R}\|$.

During the second time slot, the relay forwards the decoded signal
$\hat{s}$ through maximum ratio transmission (MRT). We assume that
the relay has partial CSI about $\textbf{h}_{R,D}$ due to channel
reciprocity impairment in TDD systems. The relation between the estimated CSI
$\hat{\textbf{h}}_{R,D}$ and the real CSI $\textbf{h}_{R,D}$ is
given by
\begin{equation}
\textbf{h}_{R,D}=\sqrt{\rho}\hat{\textbf{h}}_{R,D}+\sqrt{1-\rho}\textbf{e},\label{eqn2}
\end{equation}
where $\textbf{e}$ is the error vector with i.i.d. zero mean
and unit variance complex Gaussian entries, and is independent of
$\hat{\textbf{h}}_{R,D}$. $\rho$, scaling from $0$ to $1$, is the
correlation coefficient between $\hat{\textbf{h}}_{R,D}$ and
$\textbf{h}_{R,D}$. Thus, the received signals at the destination
and the eavesdropper are given by
\begin{equation}
y_D=\sqrt{P_R\alpha_{R,D}}\textbf{h}_{R,D}^H\textbf{r}+n_D,\label{eqn3}
\end{equation}
and
\begin{equation}
y_{E}=\sqrt{P_R\alpha_{R,E}}\textbf{h}_{R,E}^H\textbf{r}+n_{E},\label{eqn4}
\end{equation}
respectively, where $P_R$ is the transmit power of the relay,
$\textbf{r}=\textbf{v}_R\hat{s}$ is the forwarded signal with
$\textbf{v}_R=\hat{\textbf{h}}_{R,D}/\|\hat{\textbf{h}}_{R,D}\|$
being a MRT beamforming vector. $n_D$ and $n_{E}$ are additive white Gaussian noise (AWGN) samples
with unit variance at the destination and the eavesdropper,
respectively.


\section{Optimal Power Allocation}

In this section, we aim to optimize the secrecy performance through
allocating the relay power $P_R$, since it affects the signal
quality at both the destination and the eavesdropper. In what
follows, we analyze and design the power allocation schemes in the
sense of maximizing the secrecy outage capacity and minimizing the
interception probability, respectively.

\subsection{Secrecy Outage Capacity Maximization Power Allocation}
Based on the received signal in (\ref{eqn1}), when performing MRC
decoding at the relay, the channel capacity from the source to the
relay can be expressed as
\begin{equation}
C_{S,R}=W\log_2(1+\gamma_R),\label{eqn7}
\end{equation}
where $W$ is the spectral bandwidth and
$\gamma_R=P_S\alpha_{S,R}\|\textbf{h}_{S,R}\|^2$ is the
signal-to-noise ratio (SNR). Similarly, according to (\ref{eqn3})
and (\ref{eqn4}), channel capacities from the relay to the
destination and from the relay to the eavesdropper are given by
\begin{eqnarray}
C_{R,D}=W\log_2\left(1+\gamma_D\right),\label{eqn10}
\end{eqnarray}
and
\begin{eqnarray}
C_{R,E}=W\log_2\left(1+\gamma_E\right),\label{eqn11}
\end{eqnarray}
respectively, where
$\gamma_D=P_R\alpha_{R,D}\left|\textbf{h}_{R,D}^H\frac{\hat{\textbf{h}}_{R,D}}{\|\hat{\textbf{h}}_{R,D}\|}\right|^2$
and
$\gamma_E=P_R\alpha_{R,E}\left|\textbf{h}_{R,E}^H\frac{\hat{\textbf{h}}_{R,D}}{\|\hat{\textbf{h}}_{R,D}\|}\right|^2$.

Then, for the secrecy outage capacity in a DF M-MIMO secrecy
relaying system, we have the following theorem:

\emph{Theorem 1}: Given an outage probability bound by
$\varepsilon$, the secrecy outage capacity is approximated as
$C_{soc}=W\log_2\left(1+\min(P_S\alpha_{S,R}N_R,P_R\alpha_{R,D}\rho
N_R)\right)-W\log_2(1-P_R\alpha_{R,E}\ln\varepsilon)$, when the
number of relay antennas is large.

\begin{proof}
Please refer to Appendix I.
\end{proof}

Note that the secrecy outage capacity may be negative from a pure
mathematical view. Hence, it makes sense to find the condition that
the secrecy outage capacity is nonnegative.

For notional simplicity, we let $\rho \alpha_{R,D}N_R=A,
-\alpha_{R,E}\ln\varepsilon=A\cdot r_{l}, P_S\alpha_{S,R}N_R=B$,
where $r_{l}=-\frac{\alpha_{R,E}\ln\varepsilon}{\rho
\alpha_{R,D}N_R}$ is defined as the relative distance-dependent path
loss. Then, the secrecy outage capacity can be rewritten as

\begin{equation}
C_{soc}=
\begin{cases}
W\log_2(1+B)-W\log_2(1+P_RAr_l), \!\!\!\!&\mbox{$B<P_RA$}\\
W\log_2(1+P_RA)-W\log_2(1+P_RAr_l),\!\!\!\!&\mbox{$B\geq P_RA$}\nonumber
\end{cases}
\end{equation}

Observing the above secrecy outage capacity, we get the following
theorem:

\emph{Theorem 2}: If and only if $0<r_l<1$, the secrecy outage
capacity in such a DF M-MIMO secure relaying system in presence of
imperfect CSI is nonnegative.

\begin{proof}
Please refer to Appendix II.
\end{proof}

Notice that $0<r_l<1$ is the precondition for power allocation in
such a secure relaying system. Given channel conditions and outage
probability requirements, in order to fulfill $ 0<r_l<1$, the
number of antennas $N_R$ must be bigger than
$-\frac{\alpha_{R,E}\ln\varepsilon}{\rho \alpha_{R,D}}$. For an
M-MIMO relaying system, it is always possible to meet the
condition of $0<r_l<1$ by adding more antennas, which is one of its
main advantages. In what follows, we only consider the case of
$0<r_l<1$.

%
%
%
Based on Theorem 1, the secrecy outage capacity maximization power
allocation is equivalent to the following optimization problem:
\begin{eqnarray}
J_1: \max_{P_R}\quad C_{soc}\nonumber\\
s.t.\quad P_R\leq P_{\max},\label{eqn12}
\end{eqnarray}
where $P_{\max}$ is the transmit power constraint at the relay.

For the optimal solution of the above optimization problem, we have the
following theorem:

\emph{Theorem 3}: The optimal power at the relay is
$P_R^{\star}=\min\left(\frac{P_S\alpha_{S,R}}{\rho
\alpha_{R,D}},P_{\max}\right)$, and the corresponding maximum
secrecy outage capacity is
$C_{soc}^{\max}=W\log_2\left(1+\min\left(\frac{P_S\alpha_{S,R}}{\rho
\alpha_{R,D}},P_{\max}\right)A\right)-W\log_2\left(1+\min\left(\frac{P_S\alpha_{S,R}}{\rho
\alpha_{R,D}},P_{\max}\right)Ar_l\right)$.

\begin{proof}
Please refer to Appendix III.
\end{proof}

\emph{Remark}: It is found that when eavesdropper CSI is
unavailable, it is optimal for the DF secure relaying system to let
the two hops have the same channel capacity, resulting in
$P_R^{\star}=\min\left(\frac{P_S\alpha_{S,R}}{\rho
\alpha_{R,D}},P_{\max}\right)$.

\subsection{Interception Probability Minimization Power Allocation}
In this subsection, we analyze the optimal power allocation from the
perspective of minimizing the interception probability. In general,
interception probability is defined as the probability of
information leakage, namely the probability of $C_D<C_E$. Then,
interception probability is equivalent to the secrecy outage
probability when $C_{soc}=0$ in (\ref{equ5}). Thus, it can be
computed as
\begin{equation}
P_0=\exp\left(-\frac{2^{C_D/W}-1}{P_R\alpha_{R,E}}\right),\label{eqn13}
\end{equation}
where
$C_D=W\log_2\left(1+\min(P_S\alpha_{S,R}N_R,P_R\alpha_{R,D}\rho
N_R)\right)$ is the legitimate channel capacity.

Then, interception probability minimization power allocation can be
described as the following optimization problem:
\begin{eqnarray}
J_2: \min_{P_R}\quad P_0\nonumber\\
s.t.\quad P_R\leq P_{\max}.\label{eqn14}
\end{eqnarray}
Since $\exp(x)$ is a monotonously increasing function of $x$ and
$\min_{x}(-f(x))$ is equivalent to $\max_{x}(f(x))$, $J_2$ can be
transformed to the following problem:
\begin{equation}
J_3: \max_{P_R}\quad
\frac{\min(P_S\alpha_{S,R}N_R,P_R\alpha_{R,D}\rho
N_R)}{P_R\alpha_{R,E}}\nonumber
\end{equation}
\begin{equation}
s.t.\quad P_R\leq P_{\max}.\label{eqn15}
\end{equation}
By solving the above optimization problem, we have the following
theorem:

\emph{Theorem 4}: From the perspective of minimizing interception
probability, the optimal transmit power at the relay $P_R^{\star}$
belongs to a region $\left(0,\min\left(\frac{P_S\alpha_{S,R}}{\rho
\alpha_{R,D}},P_{\max}\right)\right]$, and the corresponding minimum
interception probability is
$P_0^{\min}=\exp\left(-\frac{\rho\alpha_{R,D}
N_R}{\alpha_{R,E}}\right)$.

\begin{proof}
Please refer to Appendix IV.
\end{proof}

\emph{Remarks}: It is found that the optimal power minimizing the
interception probability is not unique. However, from the perspective of maximizing
the secrecy outage capacity,
$P_R=\min\left(\frac{P_S\alpha_{S,R}}{\rho
\alpha_{R,D}},P_{\max}\right)$, namely the upper bound, is optimal.
Thus, it is better to let
$P_R^{\star}=\min\left(\frac{P_S\alpha_{S,R}}{\rho
\alpha_{R,D}},P_{\max}\right)$ in the sense of jointly optimizing
interception probability and secrecy outage capacity.

\subsection{Asymptotic Characteristic}
In the above, we prove that the optimal relay power $P_R^{\star}$ in
the sense of maximizing the secrecy outage capacity and minimizing
the interception probability is a function of the source power
$P_S$. Thus, the source power has a great impact on the secrecy
performance, as described in Theorem 3 and 4. In order to get some
clear insights, we carry out an asymptotic performance analysis with
respect to the source power.

First, for the secrecy outage capacity in a DF M-MIMO secrecy
relaying system, there are the following asymptotic
characteristics:

\emph{Proposition 1}: In the low $P_S$ regime, the optimal relay
power $P_R^*$ and maximum secrecy outage capacity $C_{soc}^{\max}$
asymptotically approach zero. In the high $P_S$ regime, the maximum
secrecy outage capacity will be saturated and is independent of
$P_S$. Furthermore, $C_{soc}^{\max}$ is an increasing function of
$P_S$.

\begin{proof}
In the low $P_S$ regime, the optimal relay power and the
corresponding secrecy outage capacity are reduced as
$P_R^{\star}=\frac{P_S\alpha_{S,R}}{\rho \alpha_{R,D}}$ and
$C_{soc}^{\max}=W\log_2\left(\frac{1+P_S\alpha_{S,R}N_R}{1+P_S\alpha_{S,R}N_Rr_l}\right)$,
respectively. Both of them asymptotically approach zero as $P_S$
tends to zero. Otherwise, in the high $P_S$ regime, the optimal
relay power is limited by $P_{\max}$, then $C_{soc}^{\max}$ is
constant. In other words, the secrecy outage capacity becomes
saturated. In addition, because of $0<r_l<1$, $C_{soc}^{\max}$ is an
increasing function of $P_S$.
\end{proof}

Second, for the interception probability, we have the following
asymptotic characteristics:

\emph{Proposition 2}: The minimum interception probability is
independent of $P_S$.

\begin{proof}
As proved in Theorem 4, although the optimal relay power is a
function of the source power, the final interception probability is
a constant independent of $P_S$ and $P_R$.
\end{proof}

\section{Simulation Results}
To examine the effectiveness of the proposed power allocation
schemes for DF M-MIMO secure relaying systems, we present several
simulation results for the following scenarios. We set $N_R=100$,
$W=10$ kHz, $\rho=0.9$ and $\varepsilon=0.01$ without extra
statements. For convenience, we normalize the path loss from the
source to the relay as $\alpha_{S,R}=1$  and use $\alpha_{R,D}$
and $\alpha_{S,E}$ to denote the relative path loss. Specifically,
$\alpha_{R,E}>\alpha_{R,D}$ means that the eavesdropper is closer
to the relay than the destination. In addition, we use
SNR$_S=10\log_{10}P_S$, SNR$_R=10\log_{10}P_R$ and
SNR$_{\max}=10\log_{10}P_{\max}$ to represent the SNR in dB at the
source, the relay and the constraint at the relay, respectively.

\begin{figure}[h] \centering
\includegraphics [width=0.45\textwidth] {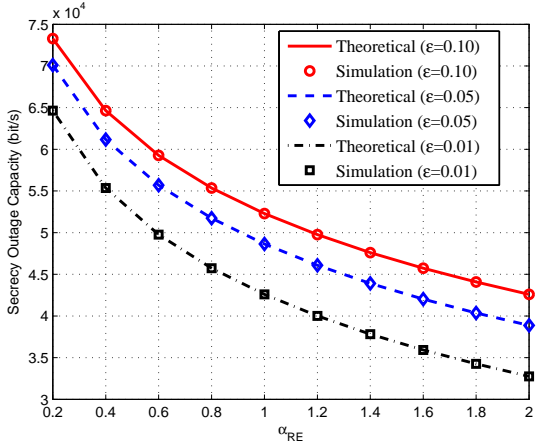}
\caption {Comparison of theoretical and simulation results with
different outage probability requirements.} \label{Fig2}
\end{figure}

First, we verify the accuracy of the theoretical expression in
Theorem 1 with SNR$_S=$ SNR$_R=10$ dB and $\alpha_{R,D}=1$. As seen
in Fig. \ref{Fig2}, the theoretical results are well consistent
with the simulations in the whole $\alpha_{R,E}$ region with
different outage probability requirements, which proves the high
accuracy of the derived results. Given an outage probability bound
by $\varepsilon$, as $\alpha_{R,E}$ increases, the secrecy outage
capacity decreases gradually. This is because the interception
ability of the eavesdropper enhances due to the short interception
distance. In addition, for a given $\alpha_{R,E}$, the secrecy
outage capacity improves with the increase of $\varepsilon$, since
the outage probability is an increasing function of the secrecy
outage capacity.

\begin{figure}[h] \centering
\includegraphics [width=0.45\textwidth] {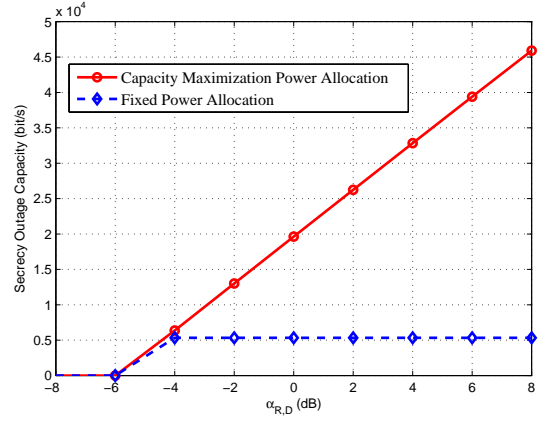}
\caption {Secrecy outage capacity comparison of different schemes.}
\label{Fig3}
\end{figure}

\begin{figure}[h] \centering
\includegraphics [width=0.45\textwidth] {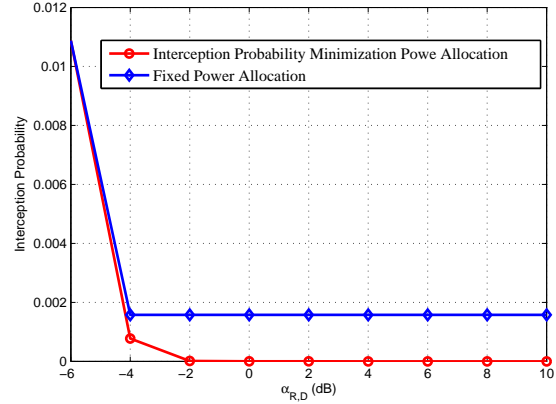}
\caption {Interception probability comparison of different schemes.}
\label{Fig4}
\end{figure}

Next, we show the performance gain of the proposed optimal power
allocation schemes compared to a fixed power allocation with
SNR$_S=10$ dB, SNR$_{\max}=15$ dB and $\alpha_{R,E}=5$. It is worth
pointing out that the fixed scheme uses a fixed power $P_R=15$ dB
regardless of channel conditions and system parameters. As seen in
Fig. \ref{Fig3}, the secrecy outage capacity maximization power
allocation scheme performs better than the fixed power allocation
scheme. Especially in the high $\alpha_{R,D}$ regime, the
performance of the proposed scheme improves sharply, while that of
the fixed allocation scheme nearly remains unchanged. This is because the
legitimate channel capacity is limited by the source-relay channel
capacity under this condition, but the fixed scheme is regardless
of $\alpha_{S,R}$ and $P_S$. In the low $\alpha_{R,D}$ regime, the
secrecy outage capacities of both schemes approach zero due to
$r_l>1$, which verifies Theorem 2 again. In terms of interception
probability, as shown in Fig. \ref{Fig4}, the proposed scheme also
outperforms the fixed power allocation scheme. Consistent with the
theoretical claims, the interception probability approaches zero
when $\alpha_{R,D}$ is large enough.

\begin{figure}[h] \centering
\includegraphics [width=0.45\textwidth] {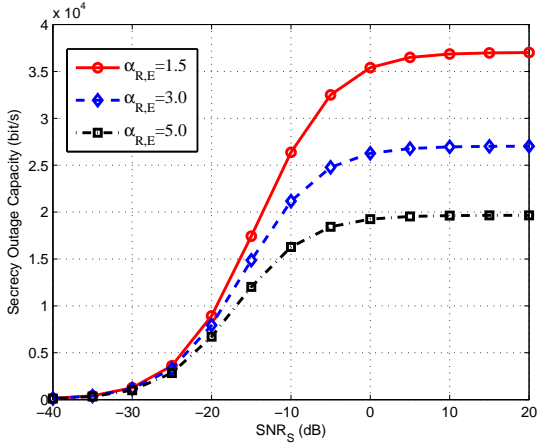}
\caption {Asymptotic secrecy outage capacity with different
$\alpha_{R,E}$.} \label{Fig5}
\end{figure}

\begin{figure}[h] \centering
\includegraphics [width=0.45\textwidth] {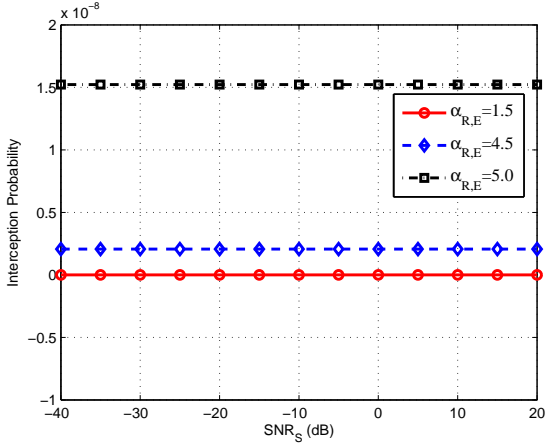}
\caption {Asymptotic interception probability with different
$\alpha_{R,E}$.} \label{Fig6}
\end{figure}

Finally, we check the asymptotic characteristics with
$\alpha_{R,D}=1$. As shown in Fig. \ref{Fig5}, as $P_S$ tends to
zero, the maximum secrecy capacities with different $\alpha_{R,E}$
approach zero. In the large $P_S$ regime, the maximum secrecy
outage capacity will be saturated, which is in agreement with Proposition 1
again. From Fig. \ref{Fig6}, it is seen that the minimum
interception probability is independent of $P_S$. Additionally,
the interception probability floor becomes higher with the increase of
$\alpha_{R,E}$.

\section{Conclusion}
In this paper, we have first presented a secrecy performance analysis for a
DF M-MIMO secure relaying system with imperfect CSI. We proved
that in order to guarantee a nonnegative secrecy outage capacity,
there is a constraint on the minimum number of antennas at the
relay. Then, by maximizing the secrecy outage capacity and
minimizing the interception probability, we proposed two optimal
relay power allocation schemes. At last, we revealed the
asymptotic characteristics of maximum secrecy outage capacity and
minimum interception probability with respect to the source power.

\begin{appendices}
\section{Proof of Theorem 1}
Based on channel capacities from the source to the relay in
(\ref{eqn7}) and from the relay to the destination in (\ref{eqn10}),
the legitimate channel capacity can be computed as
\begin{eqnarray}
C_D&=&\min(C_{S,R},C_{R,D}),\label{equ1}\\
&=&\min\Bigg(W\log_2\left(1+P_S\alpha_{S,R}\|\textbf{h}_{S,R}\|^2\right),\nonumber\\
&&W\log_2\left(1+P_R\alpha_{R,D}\Big|\textbf{h}_{R,D}^H\frac{\hat{\textbf{h}}_{R,D}}{\|\hat{\textbf{h}}_{R,D}\|}\Big|^2\right)\Bigg)\nonumber
\end{eqnarray}
\begin{eqnarray}
&=&\min\Bigg(W\log_2(1+P_S\alpha_{S,R}\|\textbf{h}_{S,R}\|^2),W\log_2\Big(1+\nonumber\\
&&P_R\alpha_{R,D}\Big|(\sqrt{\rho}\hat{\textbf{h}}_{R,D}^H+\sqrt{1-\rho}\textbf{e}^H)\frac{\hat{\textbf{h}}_{R,D}}{\|\hat{\textbf{h}}_{R,D}\|}\Big|^2\Big)\Bigg)\label{equ2}\\
&=&\min\Bigg(W\log_2(1+P_S\alpha_{S,R}\|\textbf{h}_{S,R}\|^2),W\log_2\Big(1+\nonumber\\
&&P_R\alpha_{R,D}(\rho\|\hat{\textbf{h}}_{R,D}\|^2+2\sqrt{(1-\rho)\rho}\mathcal{R}(\textbf{e}^H\hat{\textbf{h}}_{R,D})\nonumber\\
&&+(1-\rho)\|\textbf{e}^H\hat{\textbf{h}}_{R,D}\|/\|\hat{\textbf{h}}_{R,D}\|^2)\Big)\Bigg)\nonumber\\
&\approx&\min\bigg(W\log_2(1+P_S\alpha_{S,R}\|\textbf{h}_{S,R}\|^2),\nonumber\\
&&W\log_2(1+P_R\alpha_{R,D}\rho
\|\hat{\textbf{h}}_{R,D}\|^2)\bigg)\label{equ3}\\
&\approx&W\log_2(1+\min(P_S\alpha_{S,R}N_R,P_R\alpha_{R,D}\rho N_R)).\label{equ4}
\end{eqnarray}
where $\mathcal{R}(x)$ denotes the real part of $x$.
$\textbf{h}_{R,D}$ has been replaced with
$\sqrt{\rho}\hat{\textbf{h}}_{R,D}+\sqrt{1-\rho}\textbf{e}$ in
(\ref{equ2}). Eq. (\ref{equ3}) follows from the fact that
$\rho\|\hat{\textbf{h}}_{R,D}\|^2$ scales with the order
$\mathcal{O}(\rho N_R)$ as $N_R\rightarrow\infty$ while
$2\sqrt{\rho(1-\rho)}\mathcal{R}(\textbf{e}^H\hat{\textbf{h}}_{R,D})
+(1-\rho)\|\textbf{e}^H\hat{\textbf{h}}_{R,D}^H\|^2/\|\hat{\textbf{h}}_{R,D}\|^2$
scales as the order $\mathcal{O}(1)$, which is negligible. Eq.
(\ref{equ4}) holds true because of
$\lim\limits_{N_R\rightarrow\infty}\frac{\|\hat{\textbf{h}}_{R,D}\|^2}{N_R}=1$
and
$\lim\limits_{N_R\rightarrow\infty}\frac{\|\textbf{h}_{S,R}\|^2}{N_R}=1$,
namely channel hardening \cite{ChannelHardening}.

Similarly, the eavesdropper channel capacity is given by
\begin{equation}
C_E=W\log_2(1+\min(P_S\alpha_{S,R}N_R,P_R\alpha_{R,E}\Big|\textbf{h}_{R,E}^H\frac{\hat{\textbf{h}}_{R,D}}{\|\hat{\textbf{h}}_{R,D}\|}\Big|^2)).\label{app1}
\end{equation}

Then, the secrecy outage probability $\varepsilon$ with respect to a
secrecy outage capacity $C_{SOC}$ can be computed as (\ref{equ5}) at
the top of next page,
\begin{figure*}
\begin{eqnarray}
\varepsilon&=&P_r(C_{soc}>C_D-C_E)\nonumber\\
&=&P_r\left(\min\left(P_S\alpha_{S,R}N_R,P_R\alpha_{R,E}\left|\textbf{h}_{R,E}^H\frac{\hat{\textbf{h}}_{R,D}}{\|\hat{\textbf{h}}_{R,D}\|}\right|^2\right)>2^{(C_D-C_{soc})/W}-1\right)\nonumber\\
&=&P_r\left(P_S\alpha_{S,R}N_R\leq
P_R\alpha_{R,E}\left|\textbf{h}_{R,E}^H\frac{\hat{\textbf{h}}_{R,D}}{\|\hat{\textbf{h}}_{R,D}\|}\right|^2\right)P_r\left(P_S\alpha_{S,R}N_R>2^{(C_D-C_{soc})/W}-1\right)\nonumber\\
&&+P_r\left(P_S\alpha_{S,R}N_R>
P_R\alpha_{R,E}\left|\textbf{h}_{R,E}^H\frac{\hat{\textbf{h}}_{R,D}}{\|\hat{\textbf{h}}_{R,D}\|}\right|^2\right)P_r\left(P_R\alpha_{R,E}\left|\textbf{h}_{R,E}^H\frac{\hat{\textbf{h}}_{R,D}}{\|\hat{\textbf{h}}_{R,D}\|}\right|^2>2^{(C_D-C_{soc})/W}-1\right)\nonumber\\
&=&\exp\left(-\frac{P_S\alpha_{S,R}N_R}{P_R\alpha_{R,E}}\right)+\left(1-\exp\left(-\frac{P_S\alpha_{S,R}N_R}{P_R\alpha_{R,E}}\right)\right)\exp\left(-\frac{2^{(C_D-C_{soc})/W}-1}{P_R\alpha_{R,E}}\right)\label{app2}\\
&\approx&\exp\left(-\frac{2^{(C_D-C_{soc})/W}-1}{P_R\alpha_{R,E}}\right).\label{equ5}
\end{eqnarray}
\end{figure*}
where (\ref{app2}) follows from the fact that
$\Big|\textbf{h}_{R,E}^H\frac{\hat{\textbf{h}}_{R,D}}{\|\hat{\textbf{h}}_{R,D}\|}\Big|^2$
is $\chi^2$ distributed with 2 degrees of freedom, and (\ref{equ5})
holds true since
$\exp\left(-\frac{P_S\alpha_{S,R}N_R}{P_R\alpha_{R,E}}\right)$
approaches zero when $N_R$ is sufficient large. Based on
(\ref{equ5}), it is easy to get the Theorem 1.

\section{Proof of Theorem 2}
Based on the secrecy outage capacity in Theorem 1, when
$P_R\geq\frac{P_S\alpha_{S,R}}{\rho \alpha_{R,D}}$, we have
$C_{soc}=W\log_2(1+P_S\alpha_{S,R}N_R)-W\log_2(1+P_R\rho
\alpha_{R,D}N_Rr_l)$. To guarantee $C_{soc}\geq0$, the following
condition $P_S\alpha_{S,R}N_R\geq P_R\rho \alpha_{R,D}N_Rr_l$ must
be fulfilled, which is equivalent to $0<r_l<1$ in the case of
$P_R>\frac{P_S\alpha_{S,R}}{\rho \alpha_{R,D}}$.

Otherwise, when $P_R\leq \frac{P_S\alpha_{S,R}}{\rho \alpha_{R,D}}$,
the secrecy outage capacity is changed as $C_{soc}=W\log_2(1+P_R\rho
\alpha_{R,D}N_R)-W\log_2\left(1+P_R\rho\alpha_{R,D}N_Rr_l\right)$.
Only when $0<r_l<1$, $C_{soc}$ is nonnegative.

Above all, $0<r_l<1$ or $N_R>-\frac{\alpha_{R,E}\ln\varepsilon}{\rho
\alpha_{R,D}}$ is the precondition that the nonnegative secrecy
outage capacity exists. Therefore, we get the Theorem 2.

\section{proof of Theorem 3}
According to the Theorem 1, when $P_R\geq\frac{P_S\alpha_{S,R}}{\rho
\alpha_{R,D}}$, the secrecy outage capacity
$C_{soc}=W\log_2(1+P_S\alpha_{S,R}N_R)-W\log_2(1+P_R\rho
\alpha_{R,D}N_Rr_l)$ is maximized when
$P_R=\frac{P_S\alpha_{S,R}}{\rho \alpha_{R,D}}$, since $C_{soc}$ is
a monotonously decreasing function of $P_R$.

When $P_R\leq \frac{P_S\alpha_{S,R}}{\rho \alpha_{R,D}}$, the
secrecy outage capacity $C_{soc}=W\log_2(1+P_R\rho
\alpha_{R,D}N_R)-W\log_2(1+P_R\rho \alpha_{R,D}N_Rr_l)$ is an
increasing function of $P_R$ under the condition $0<r_l<1$. Thus,
$P_R=\frac{P_S\alpha_{S,R}}{\rho \alpha_{R,D}}$ is the optimal power
at the relay.

Considering the constraint on the transmit power $P_{\max}$ at the
relay, the optimal power at the relay is
$P_R^{\star}=\min\left(\frac{P_S\alpha_{S,R}}{\rho
\alpha_{R,D}},P_{\max}\right)$. Furthermore, by substituting
$P_R^{\star}=\min\left(\frac{P_S\alpha_{S,R}}{\rho
\alpha_{R,D}},P_{\max}\right)$ into the expression of secrecy outage
capacity, we can obtain the maximum secrecy outage capacity as shown
in Theorem 3.

\section{proof of Theorem 4}
First, when $P_R\leq\frac{P_S\alpha_{S,R}}{\rho \alpha_{R,D}}$, the
optimization problem $J_3$ is equivalent to
\begin{equation}
G_1 : \max_{P_R}\quad \frac{\alpha_{R,D}\rho
N_R}{\alpha_{R,E}}\nonumber
\end{equation}
\begin{equation}
s.t. \quad P_R\leq \min\left(\frac{P_S\alpha_{S,R}}{\rho
\alpha_{R,D}},P_{\max}\right).\label{equ9}
\end{equation}
Interestingly, it is found that the objective function $\frac{\rho
\alpha_{R,D}N_R}{\alpha_{R,E}}$ is independent of $P_R$. Hence, the
optimal solution of $G_1$ can be an arbitrary value belonging to
$\left(0,\min\left(\frac{P_S\alpha_{S,R}}{\rho
\alpha_{R,D}},P_{\max}\right)\right]$.

Second, when $P_R\geq\frac{P_S\alpha_{S,R}}{\rho \alpha_{R,D}}$, the
optimization problem $J_3$ is reduced as
\begin{equation}
G_2 : \max_{P_R}
\quad\frac{P_S\alpha_{S,R}N_R}{P_R\alpha_{R,E}}\nonumber
\end{equation}
\begin{equation}
s.t. \quad \frac{P_S\alpha_{S,R}}{\rho \alpha_{R,D}}\leq P_R\leq
P_{\max}.\label{equ10}
\end{equation}
The optimal solution of $G_2$ is $\frac{P_S\alpha_{S,R}}{\rho
\alpha_{R,D}}$, since the objective function is a decreasing
function of $P_R$.

Thus, the optimal transmit power at the relay is
$P_R^{\star}=\left(0,\min\left(\frac{P_S\alpha_{S,R}}{\rho
\alpha_{R,D}},P_{\max}\right)\right]$. Hence, we get the Theorem 4.
\end{appendices}

\end{document}